# Scattering Phases and Density of States
# for Exterior Domains

*J.-P. Eckmann*[1,2] *and C.-A. Pillet*[1]

[1]Dépt. de Physique Théorique, Université de Genève, CH-1211 Genève 4, Switzerland
[2]Section de Mathématiques, Université de Genève, CH-1211 Genève 4, Switzerland

**Abstract.** For a bounded open domain $\Omega \in \mathbf{R}^2$ with connected complement and piecewise smooth boundary, we consider the Dirichlet Laplacian $-\Delta_\Omega$ on $\Omega$ and the S-matrix on the complement $\Omega^c$. Using the restriction $A_E$ of $(-\Delta - E)^{-1}$ to the boundary of $\Omega$, we establish that $A_{E_0}^{-1/2} A_E A_{E_0}^{-1/2} - 1$ is trace class when $E_0$ is negative and give bounds on the energy dependence of this difference. This allows for precise bounds on the total scattering phase, the definition of a $\zeta$-function, and a Krein spectral formula, which improve similar results found in the literature.

# 1. Introduction

We consider bounded domains $\Omega$ in $\mathbf{R}^2$ and study the Dirichlet Laplacian $\Delta_\Omega$ in $\Omega$ as well as the scattering matrix (also with Dirichlet condition) in the complement $\Omega^c$. In the paper [EP] we have begun this study by describing a "spectral duality" between the eigenvalues of the Laplacian and the scattering phases of the S-matrix restricted to an energy shell. In this paper, we will go further and show how $\zeta$-functions for this problem can be derived. This allows to establish a relatively precise relation between the eigenvalues of the Laplacian and the phase shifts, which leads to improvements of the results of [JK, M] and extensions to more complicated domains. Similar such relations have been studied numerically and in perturbation theory in [DS]. We also establish a Krein trace formula.

To set the stage, we define "standard" domains $\Omega$ and we then assume throughout that $\Omega$ is a standard domain.

**Definition.** A domain $\Omega \subset \mathbf{R}^2$ is called a *standard domain* if it has the following properties: $\Omega$ is a bounded, open set, whose boundary is piecewise $\mathcal{C}^2$ with a finite number of pieces. Furthermore, the angles at the corners are required to be non-degenerate, i.e., different from 0 and $2\pi$. Finally, the complement of the closure of $\Omega$ is connected.

Thus, a standard domain is for example a union of squares and circles, and it need not be connected nor convex. In order to simplify the notation, we shall only consider connected domains, but the proofs carry through without problems for the general case, by replacing the function spaces by direct sums of the spaces for each piece of $\Omega$. We also assume for simplicity that the perimeter of $\Omega$ has length $2\pi$. Throughout, $x$ denotes the $2\pi$-periodic map

$$x : S^1 \to \Gamma = \partial\Omega \subset \mathbf{R}^2 \,,$$

which maps (isometrically) the arclength to the boundary.

We next define the central object of study: The boundary Green's function $A_E$. With a slight change of notation from [EP] we let $A_E$ denote the integral operator (and the integral kernel) of the Green's function on the boundary:

$$\begin{aligned} A_E &: S^1 \times S^1 \to \mathbf{C} \,, \\ A_E(s,s') &= G_E(|x(s) - x(s')|) \,. \end{aligned} \quad (1.1)$$

Here, $G_E$ is the Green's function of the "free" Laplacian on $\mathbf{R}^2$:

$$G_E = \frac{1}{-\Delta - E} \,,$$

whose integral kernel is

$$G_E(x) = \tfrac{i}{4} H_0^+(\sqrt{E}|x|) = \tfrac{i}{4} J_0(\sqrt{E}|x|) - \tfrac{1}{4} Y_0(\sqrt{E}|x|) \,. \quad (1.2)$$

The functions $J_0$, $Y_0$ and $H_0^+$ are the Bessel and Hankel functions [AS]. Some care is needed with the square root:

**Definition.** We let $\mathcal{E}$ denote $\{E : E \in \mathbf{C}, E \notin \mathbf{R}^+\}$. We denote by $\sqrt{E}$ the root of $E$ with the following determination: For $E > 0$, $\sqrt{E + i0}$ is the positive root, and then $\sqrt{E - i0}$ is the negative one. So the square root maps $\mathcal{E}$ to the upper halfplane. We also denote by $\mathcal{R}$ the Riemann surface associated with the logarithm.

**Remark.** The operator $A_E$ is analytic in $\mathcal{R}$.

In [EP], properties of $A_E$ were studied which allowed us to prove several results about the Laplacian and the S-matrix, when restricted to the energy shell $E$. The definition of $S_E$ will be given below, but see [EP] for a rederivation from first principles of scattering theory. We showed

**Theorem 1.1.** *The number $E_0 > 0$ is an $m$-fold eigenvalue of $-\Delta_\Omega$ if and only if exactly $m$ scattering phases $\vartheta_j(E)$ of the S-matrix $S_E$ tend to $\pi$ as $E \uparrow E_0$.*

In the current paper, we improve our control over the operator $A_E$ and use it to define $\zeta$-functions, and Krein formulas.

Our first main result is the Structure Theorem II which says that

$$T_E = A_{-1}^{-1/2} A_E A_{-1}^{-1/2} - 1$$

is an operator in the Birman-Solomiak class $\mathcal{S}_{2/3}$ and satisfies the bound

$$\langle T_E \rangle_{2/3} \leq K |E|^{3/4} \log|E| \quad \text{as} \quad |E| \to \infty . \tag{1.3}$$

Similar bounds are known in potential scattering, but our bound is new for the problem of obstacle scattering, and stronger than earlier bounds in the literature. The definition of $A_E$ and the bound Eq.(1.3) allow for a very simple definition of a $\zeta$-function in the present case of obstacle scattering, i.e., hard-core potentials, and an identity for the determinant of the S-matrix (Theorem 3.1):

$$\zeta(E) = \det(1 + T_E) , \quad S_E = \bar{\zeta}(E + i0)/\zeta(E + i0) .$$

Note that these quantities are expressed by operators on the boundary of the obstacle. One can use all these estimates to improve the results of earlier papers [JK, S, MR], (Theorem 3.2), which gives pointwise bounds on the total phase shift $\Theta$ and number of bound states for a very general class of domains:

$$|\Theta(E) - \pi N(E)| \leq |E|^{1/2} \log|E|, \quad \text{as} \quad |E| \to \infty . \tag{1.4}$$

The earlier estimates where for the integral over $E$ of this quantity (for obstacle scattering). Finally, in the same vein, the Krein formula can be derived in the same framework.

## 2. Structure Theorems

In this section, we derive very detailed representations of the operators $A_E$ which we call structure theorems. They tell us that properly regularized versions of $A_E$ are proportional to $(1+$ trace class$)$. Denoting by $\partial_s$ the derivative with respect to arclength, and setting $\Lambda = (1 + (i\partial_s)^2)^{1/2}$, the regularizations which we consider are:

$\Lambda A_E$, and

$A_{E_0}^{-1/2} A_E A_{E_0}^{-1/2}$, where $E_0$ is an arbitrary negative constant.

**Notation.** If $C$ is a compact operator, we let $s_n(C)$, $n = 1, 2, \ldots$ be the $n^{\text{th}}$ eigenvalue of $(C^*C)^{1/2}$ (in decreasing order). We define for $1 \leq p < \infty$, and for $p = \infty$ the usual norms

$$\|C\|_p = \left(\sum_{n=1}^{\infty} s_n(C)^p\right)^{1/p},$$

$$\|C\|_\infty = \sup_n s_n(C) = s_1(C).$$

We also need the weak Schatten classes[S, BS]: We let

$$\langle C \rangle_p = \sup_n n^{1/p} s_n(C), \quad \langle C \rangle_\infty = \|C\|_\infty. \tag{2.1}$$

The class $\Sigma_p$ of operators $C$ with finite $\langle C \rangle_p$ is a complete topological vector space for $p > 0$ and is normable for $p > 1$.

**Remark.** We shall need the inequalities:

$$\langle C \rangle_p \leq \|C\|_p, \quad \|C\|_1 \leq \langle C \rangle_{2/3}, \tag{2.2}$$

which are obvious from the definition, and the more subtle one, see [BS],

$$\langle C_1 C_2 \rangle_p \leq 2^{1/p} \langle C_1 \rangle_q \langle C_2 \rangle_r, \tag{2.3}$$

which holds for all $q > 0, r > 0, p^{-1} = q^{-1} + r^{-1}$.

**Notations.**
— The bounds which will be given below are of the form $|E|^\alpha \log |E|$. We shall use the shorthand notation
$$|E|^{\alpha+} = |E|^\alpha \log |E|.$$
— Constants, such as $K$, which are used in bounds can change their meaning from one equation to the next.
— We let $P_0$ denote the orthogonal projection onto the constant functions in $L^2(S^1)$.

Our main technical result is the

**Structure Theorem I.** *For $E \in \mathcal{R}$ the operator $\Lambda A_E$ has the following representation:*

$$\Lambda A_E = \tfrac{1}{2} + B + H + T_E^{(1)}, \tag{2.4}$$

where $B$ is bounded, and of norm $\|B\| < \frac{1}{2}$, $H$ is Hilbert-Schmidt, and $T_E^{(1)}$ is trace class. There is a constant $K$ so that for $E \in \mathcal{E}$ one has the bounds

$$\langle T_E^{(1)} + \tfrac{1}{2} P_0 \log E \rangle_{2/3} \leq K|E|^{3/4+} ,$$
$$\|T_E^{(1)} \Lambda + \tfrac{1}{2} P_0 \log E\|_2 \leq K|E|^{3/4+} . \quad (2.5)$$

The proof will be given in Sect.5.

**Remarks.**
1) The bound (2.5) suggests that the $n^{\text{th}}$ eigenvalue of $T_E^{(1)}$ is about $E^{3/4} n^{-3/2}$ (for large $E$).
2) The choice of $H$ is somewhat arbitrary, since we can add to it part of the trace class operator without changing the statement of the theorem.
3) If $\Gamma = \partial\Omega$ is $\mathcal{C}^2$ (i.e., if there are no corners) then one can choose $B = 0$.
4) The improvement of this result over the structure theorem in [EP, Theorem 4.1, Eq.(4.25)] is the observation that those parts of $\Lambda A_E$ which are not trace class do not depend on the energy. The bounds (2.5) are also new.

Based on 4), we now proceed as follows: Let $E_0$ be an arbitrary negative constant, which we fix throughout the remainder of the paper. Since $-\Delta$ has spectrum in $\mathbf{R}^+$ one can check from the definition of $A_E$ that $A_{E_0}$ is invertible. Therefore the following statement makes sense:

**Structure Theorem II.** *Let $E \in \mathcal{R}$ and $E_0 < 0$. Then one has the representation*

$$A_{E_0}^{-1/2} A_E A_{E_0}^{-1/2} = 1 + T_E , \quad (2.6)$$

*where $T_E$ is trace class. For $E \in \mathcal{E}$, and $|E| > 1$, one has the bound*

$$\langle T_E \rangle_{2/3} \leq K|E|^{3/4+} . \quad (2.7)$$

*Furthermore, there are a rank one orthogonal projection $P$ and a constant $C > 0$ such that for $E \in \mathcal{E}$, and $|E| < 1$,*

$$\langle T_E - CP \log(E/E_0) \rangle_{2/3} \leq K|E|^{3/4+} . \quad (2.8)$$

*Finally, $T_E^* = T_{\bar{E}}$, and if $\operatorname{Im} E > 0$, then $\operatorname{Im} T_E > 0$*

**Proof.** We start by deriving a few consequences of the Structure Theorem I. Note first that $\Lambda A_{E_0}$ is of the form

$$\Lambda A_{E_0} = \tfrac{1}{2} + B + H + T_{E_0}^{(1)} ,$$

and 0 is not in its spectrum since $E_0 < 0$. Therefore,

$$\|\Lambda A_{E_0}\|_\infty < \infty , \quad \|(\Lambda A_{E_0})^{-1}\|_\infty < \infty .$$

Throughout, we shall need the bound

$$\langle \Lambda^{-1} \rangle_1 < \infty , \quad (2.9)$$

which follows by observing that the spectrum of $\Lambda$ is $\{\sqrt{1+n^2}\}_{n\in\mathbf{Z}}$. We next bound $\langle A_{E_0}^{1/2}\rangle_2$. Since $A_{E_0}$ is positive, we have

$$\langle A_{E_0}^{1/2}\rangle_2^2 = \sup_n n\, s_n(A_{E_0}^{1/2})^2 = \sup_n n\, s_n(A_{E_0}) = \langle A_{E_0}\rangle_1$$
$$= \langle \Lambda^{-1}\Lambda A_{E_0}\rangle_1 \leq \langle \Lambda^{-1}\rangle_1 \langle \Lambda A_{E_0}\rangle_\infty \leq \mathrm{const.}\|\Lambda A_{E_0}\|_\infty\ .$$

Using now the Structure Theorem I, we see that for $|E|>1$,

$$\langle \Lambda(A_E - A_{E_0})\Lambda\rangle_2 = \langle (T_E^{(1)} - T_{E_0}^{(1)})\Lambda\rangle_2 \leq \mathrm{const.}\left(\langle T_E^{(1)}\Lambda\rangle_2 + \langle T_{E_0}^{(1)}\Lambda\rangle_2\right) \leq \mathrm{const.}|E|^{3/4+}\ .$$

It is straightforward that

$$T_E = A_{E_0}^{-1/2} A_E A_{E_0}^{-1/2} - 1 = A_{E_0}^{1/2}\left(\Lambda A_{E_0}\right)^{-1}\cdot\Lambda(A_E - A_{E_0})\Lambda\cdot\left(A_{E_0}\Lambda\right)^{-1} A_{E_0}^{1/2}\ .$$

Using Eq.(2.3), we obtain, for $|E|>1$,

$$\langle T_E\rangle_{2/3} \leq \mathrm{const.}\langle A_{E_0}^{1/2}\rangle_2 \langle\left(\Lambda A_{E_0}\right)^{-1}\rangle_\infty \cdot \langle \Lambda(A_E - A_{E_0})\Lambda\rangle_2 \cdot \langle\left(A_{E_0}\Lambda\right)^{-1}\rangle_\infty \langle A_{E_0}^{1/2}\rangle_2\ .$$

Substituting the previous bounds and observing that $A_{E_0}\Lambda = (\Lambda A_{E_0})^*$, we obtain the inequality (2.7). Defining $P_1 = A_{E_0}^{-1/2} P_0 A_{E_0}^{-1/2}$ and $P = P_1/\|P_1\|$, one obtains the bound (2.8). Since $A_{E_0}$ is selfadjoint, and $G(\bar E) = (G(E))^*$, the last assertions follow because $G(E)$ is a Herglotz function. The proof of the Structure Theorem II is complete.

## 3. The $\zeta$-function

By the Structure Theorem II we can define the analytic function of $E \in \mathcal{R}$:

$$\zeta(E) = \det\left(A_{E_0}^{-1/2} A_E A_{E_0}^{-1/2}\right) = \exp\,\mathrm{Tr}\,\log(1 + T_E)\ . \tag{3.1}$$

From the Herglotz property we conclude immediately that $\arg\zeta(E+i0) = 0$ if $E < 0$, and is positive if $E \geq 0$. We have the following

**Theorem 3.1.** *The determinant of $S_E$ is given, for $E>0$, by*

$$\det(S_E) = \frac{\bar\zeta(E+i0)}{\zeta(E+i0)}\ . \tag{3.2}$$

The proof will be given at the end of this section.

**Definition.** We define the *total scattering phase* $\Theta(E)$ by the identity

$$e^{-2i\Theta(E)} = \det(S_E)\ .$$

Since $S_E$ is analytic (as can be seen, e.g. from Eq.(3.7) below), $\Theta$ can be chosen continuous. There is an overall indeterminacy of $n\pi$ which we eliminate by requiring $\Theta(0) = 0$. This choice is possible because $\det(S_{E=0}) = 1$. Indeed, Eq.(2.8) implies $\lim_{E \to 0} \text{Im}\,\zeta(E+i0) = 0$, from which $\det(S_{E=0}) = 1$ follows.

We next define $N(E) : \mathbf{R}^+ \to \mathbf{Z}$ as the integrated density of states of $-\Delta_\Omega$, i.e., the number of eigenvalues of $-\Delta_\Omega$ below $E$. Then we have the important identity:

$$\Theta(E) = \pi N(E) + \text{Im}\,\log\,\zeta(E+i0)\,. \tag{3.3}$$

This can be seen as follows: It is a well-known fact from potential theory—and reproved in [EP, Lemma 5.5]—that $A_E$ has an $m$-fold eigenvalue 0 if and only if $-\Delta_\Omega$ has an $m$-fold eigenvalue equal to $E$. Therefore, the same is true for $A_{E_0}^{-1/2} A_E A_{E_0}^{-1/2}$, and thus, by Eq.(3.1), the quantity $\text{Im}\,\log\,\zeta(E+i0)$ jumps by $-m\pi$ at each such eigenvalue. It is continuous elsewhere, since in fact $T_E$ is a real analytic function of $E$. It follows that $\pi N(E) + \text{Im}\,\log\,\zeta(E+i0)$ is also continuous. Thus, Eq.(3.3) holds.

We obtain the following important bound, which improves [JK, S, MR]:

**Theorem 3.2.** *Let $\Omega$ be a standard domain and $E > 1$. There is a $K$ such that*

$$0 \leq \Theta(E) - \pi N(E) \leq KE^{1/2+}\,. \tag{3.4}$$

**Remark.** Note that $N(E)$ typically grows like $\mathcal{O}(E)$ so that the bound says that the phase shift and the integrated density of states are "comparable" in this case. Of course, in the case of (smooth) potential scattering, one has additional information about the phase shift, so that inequalities like Eq.(3.4) give direct information on $N(E)$. This is not the case for the much more singular problem considered here, where the resonances can accumulate near the real axis from below as $E \uparrow \infty$. Furthermore, extending slightly [MR], or from numerical experiments [U], one can see that if $\Omega$ is a circle of radius $R$, then an averaged version $\bar{N}$ of $N$ and the phase shift satisfy

$$\bar{N}(E) = ER^2/4 - \sqrt{E}R/2 + \mathcal{O}(1)\,,$$
$$\Theta(E)/\pi = ER^2/4 + \sqrt{E}R/2 + 1/6 + \mathcal{O}(E^{-1/2})\,.$$

Therefore the difference in Eq.(3.4) cannot be smaller than $\mathcal{O}(E^{1/2})$.

**Proof of Theorem 3.2.** Starting with Eq.(3.3), we see that $|\Theta(E) - \pi N(E)| = |\text{Im}\,\log\,\zeta(E+i0)|$. From Eq.(3.1), we deduce that

$$|\text{Im}\,\log\,\zeta(E+i0)| = |\text{Im}\,\log\det(1 + T_{E+i0})|\,. \tag{3.5}$$

Since $T_E$ has the Herglotz property, we can apply the inequality of Sobolev [S, Lemma 2.2] to obtain

$$|\text{Im}\,\log\det(1 + T_{E+i0})| \leq \text{const.} < T_{E+i0} >_{2/3}^{2/3}\,. \tag{3.6}$$

Substituting the bounds of the Structure Theorem II, the assertion of Theorem 3.2 follows.

We can draw another nice conclusion from the Structure Theorems:

**Proposition 3.3.** *Let $\Omega$ be a standard domain. There is a constant $D$ such that the multiplicity of an eigenvalue $E$ of $-\Delta_\Omega$ is bounded by $DE^{1/2+}$.*

**Proof.** The multiplicity of the eigenvalues less than 1 is bounded. It suffices thus to consider $E > 1$. We recall the result [EP] that $-\Delta_\Omega$ has an eigenvalue $E$ of multiplicity $m$ if and only if $A_E$ has an eigenvalue 0 of multiplicity $m$. Since $A_{E_0}$ is invertible, $A_{E_0}^{-1/2} A_E A_{E_0}^{-1/2}$ has an eigenvalue 0 of multiplicity $m$ in this case. But $A_{E_0}^{-1/2} A_E A_{E_0}^{-1/2} = 1 + T_E$, so that $T_E$ has an eigenvalue $-1$ of multiplicity $m$. Since we have shown in Eq.(2.7) that $\langle T_E \rangle_{2/3} \leq \mathcal{O}(E^{3/4+})$ for $E \in \mathcal{E}$ the assertion follows by observing that $m^{3/2} \cdot |-1| \leq \langle T_E \rangle_{2/3} = \mathcal{O}(E^{3/4+})$.

**Proof of Theorem 3.1.** Our starting point is the following representation of $S_E$ [EP, Eq.(3.15)]:

$$S_E = 1 - 2i\mathbf{J}_{E+i0}^{1/2} A_{E+i0}^{-1} \mathbf{J}_{E+i0}^{1/2}, \qquad (3.7)$$

which holds for $E > 0$. The operator $\mathbf{J}_E$ and its counterpart $\mathbf{Y}_E$ are defined through their integral kernels

$$\mathbf{J}_z(s,s') = \tfrac{1}{4} J_0(\sqrt{z}|x(s) - x(s')|),$$
$$\mathbf{Y}_z(s,s') = -\tfrac{1}{4} Y_0(\sqrt{z}|x(s) - x(s')|),$$

so that $A_z = \mathbf{Y}_z + i\mathbf{J}_z$. The following facts are straightforward consequences of the properties of the $J_0$ and $Y_0$ functions [AS, §9]:

- $J_0$ is entire, and $J_0(\bar{w}) = \bar{J}_0(w)$. Therefore, $J_0(e^{i\pi m} w) = J_0(w)$.
- $Y_0$ has a branch point at $w = 0$ (which we lift by putting a branch cut on $\mathbf{R}^-$), and $Y_0(\bar{w}) = \bar{Y}_0(w)$. Finally, $Y_0(e^{i\pi m} w) = Y_0(w) + 2im J_0(w)$.

Using the determination for $\sqrt{w}$ as defined for $\sqrt{E}$, (i.e., $\sqrt{k^2 + i0} = |k|$) the above identities imply in terms of the operators:

$$\begin{aligned}
\mathbf{J}_z^* &= \mathbf{J}_{\bar{z}}, & \mathbf{J}_{e^{2\pi i} z} &= \mathbf{J}_z, \\
\mathbf{Y}_z^* &= \mathbf{Y}_{\bar{z}} + 2i\mathbf{J}_{\bar{z}}, & \mathbf{Y}_{e^{2\pi i} z} &= \mathbf{Y}_z - 2i\mathbf{J}_z, \\
A_z^* &= \mathbf{Y}_z^* - i\mathbf{J}_z^* = A_{\bar{z}}, & A_{e^{2\pi i} z} &= \mathbf{Y}_z - i\mathbf{J}_z.
\end{aligned} \qquad (3.8)$$

Consider now

$$C_z = A_{E_0}^{-1/2} A_z A_{E_0}^{-1/2}, \qquad (3.9)$$

so that by the Structure Theorem II,

$$C_z = 1 + T_z.$$

We find from Eq.(3.7) that for $z \notin \sigma(-\Delta_\Omega)$, one has

$$\begin{aligned}
\det(S_z) &= \det\left(1 - 2i\mathbf{J}_z^{1/2} A_z^{-1} \mathbf{J}_z^{1/2}\right) \\
&= \det\left(1 - 2iA_z^{-1} \mathbf{J}_z\right) \\
&= \det\left(A_z^{-1}(A_z - 2i\mathbf{J}_z)\right) \\
&= \det\left(A_z^{-1}(\mathbf{Y}_z - i\mathbf{J}_z)\right) \\
&= \det\left(A_z^{-1} A_{e^{2\pi i} z}\right) \\
&= \det\left(C_z^{-1} C_{e^{2\pi i} z}\right),
\end{aligned}$$

where the last equality holds by Eq.(3.9). Using Eq.(3.8), we see that $A_z^* = A_{\bar z}$ implies $T_z^* = T_{\bar z}$. Therefore, since

$$\det(S_z) = \frac{\det(1 + T_{e^{2\pi i z}})}{\det(1 + T_z)} ,$$

the assertion Eq.(3.2) follows from

$$\zeta(\bar z) = \det(1 + iT_{\bar z}) = \det((1 + iT_z)^*) = \bar\zeta(z) .$$

The proof of Theorem 3.1 is complete.

## 4. The Krein trace formula

We consider here the "free" Hamiltonian $H_0 = -\Delta$ and the "interacting" Hamiltonian $H = -\Delta_\Omega \oplus -\Delta_{\Omega^c}$. Then one has the

**Theorem 4.1.** *For every $F \in \mathcal{S}(\mathbf{R})$ with support in $\{E : E > 0\}$ one has the identity*

$$\operatorname{Tr}\bigl(F(H) - F(H_0)\bigr) = \sum_n F(\lambda_n) + \frac{1}{2\pi i} \int dE\, F(E)\operatorname{Tr}(S_E^* \partial_E S_E) , \qquad (4.1)$$

*where the $\lambda_n$ are the eigenvalues of $-\Delta_\Omega$ and $S_E$ denotes the on-shell S-matrix.*

**Remark.** The condition on $F$ given above is too strong. One can for example relax it along the lines of [Y, Theorem 8.3.3]. In another direction, probably more useful for applications, it seems that $E^4 F'' + E^3 F' \in L^2$ is a sufficient condition (at $E$ near $\infty$).

**Proof.** The proof is an application of the usual Krein trace formula. All we have to show is essentially that $(H - z)^{-1} - (H_0 - z)^{-1}$ is trace class, and then perform a few changes of variables. It follows from the definition of $H$, $H_0$, that (with a slight change of notation from the other sections of this paper),

$$G_0(z) \equiv (H_0 - z)^{-1} , \qquad (4.2)$$
$$G(z) \equiv (H - z)^{-1} = G_0(z) - G_0(z)\gamma^* A_z^{-1} \gamma G_0(z) , \qquad (4.3)$$

which we proved in [EP, Eq.(5.10)]. Here, $\gamma$ is the operator which restricts a function on $\mathbf{R}^2$ to the boundary $\Gamma = \partial\Omega$. Let $E_0 < 0$ and define

$$V = G(E_0) - G_0(E_0) = -G_0(E_0)\gamma^* A_{E_0}^{-1} \gamma G_0(E_0) . \qquad (4.4)$$

One can relate this "Hamiltonian" formalism with the $\zeta$-function we considered above:

**Lemma 4.2.** *One has the bound*
$$\|V\|_1 < \infty , \qquad (4.5)$$

and (for $E \in \mathcal{E}$) the identity

$$\zeta(E) = \det\left(1 + V \cdot \left(G_0(E_0) - (E - E_0)^{-1}\right)^{-1}\right). \tag{4.6}$$

The proof will be given at the end of this section. We define, for $\lambda \in \mathbf{R}$,

$$\xi(\lambda) = \pi^{-1} \arg \det\left(1 + V\left(G_0(E_0) - \lambda - i0\right)^{-1}\right). \tag{4.7}$$

By Eq.(4.5) and Krein's theorem [Y,Theorem 8.3.3], this definition makes sense and one has furthermore for all $f$ for which $f'$ is the Fourier transform of a finite (complex) measure, the identity

$$\mathrm{Tr}\big(f(G(E_0)) - f(G_0(E_0))\big) = \int d\lambda\, \xi(\lambda) f'(\lambda). \tag{4.8}$$

Assume now $F$ satisfies the conditions of Theorem 4.1. If we define $f$ by $f\big((E - E_0)^{-1}\big) = F(E)$, then we can apply Eq.(4.8) for this $f$. Defining

$$\eta(E) \equiv -\xi((E - E_0)^{-1}),$$

we get:

$$\begin{aligned}
\mathrm{Tr}\big(F(H) - F(H_0)\big) &= \mathrm{Tr}\big(f(G(E_0)) - f(G_0(E_0))\big) \\
&= \int d\lambda\, \xi(\lambda) f'(\lambda) \\
&= \int \frac{dE}{(E - E_0)^2}\, \xi\big((E - E_0)^{-1}\big)\, f'\big((E - E_0)^{-1}\big) \\
&= \int dE\, \eta(E) F'(E).
\end{aligned} \tag{4.9}$$

By Eqs.(4.6) and (4.7), we find that

$$\xi\big((E - E_0)^{-1}\big) = -\pi^{-1} \arg \zeta(E + i0).$$

We next note that from its definition, $\Theta'(E) = -(2i)^{-1} \mathrm{Tr}(S_E^* \partial_E S_E)$. Therefore, when $E > 0$,

$$\begin{aligned}
-\eta'(E) &= \partial_E \xi\big((E - E_0)^{-1}\big) = -\pi^{-1} (\arg \zeta)'(E + i0) \\
&= N'(E) - \pi^{-1} \Theta'(E) = \sum_n \delta(E - \lambda_n) + (2\pi i)^{-1} \mathrm{Tr}(S_E^* \partial_E S_E).
\end{aligned}$$

Since we assumed $F \in \mathcal{S}$ with support in $E > 0$, we can integrate by parts in Eq.(4.9) and obtain Eq.(4.1). The proof of Theorem 4.1 is complete.

**Proof of Lemma 4.2.** We show first that $V$ is trace class. Using Eq.(4.4) we write $V = -L^* L$, where $L = A_{E_0}^{-1/2} \gamma G_0(E_0)$. We shall bound $\|V\|_1$ by showing that $\|V\|_1 = \|L\|_2^2 = \|L^*\|_2^2$ is finite. By the Structure Theorem II, we know that $T_z$ analytic and trace class, and therefore its derivative is also trace class. The resolvent identity thus implies

$$\partial_z T_z = \partial_z \left( A_{E_0}^{-1/2} A_z A_{E_0}^{-1/2} - 1 \right)$$
$$= \partial_z A_{E_0}^{-1/2} \gamma G_0(z) \gamma^* A_{E_0}^{-1/2} = A_{E_0}^{-1/2} \gamma G_0(z)^2 \gamma^* A_{E_0}^{-1/2} \ .$$

Therefore,

$$\|L^*\|_2^2 = \mathrm{Tr}(A_{E_0}^{-1/2} \gamma G_0(E_0)^2 \gamma^* A_{E_0}^{-1/2}) = \left\| \partial_z T_z \big|_{z=E_0} \right\|_1 < \infty \ ,$$

and this proves Eq.(4.5).

We next note the resolvent identity

$$G_0(E_0) - (E - E_0)^{-1} = \frac{1}{H_0 - E_0} - \frac{1}{E - E_0} = -\frac{1}{E - E_0} \frac{H_0 - E}{H_0 - E_0} \ . \qquad (4.10)$$

By the Structure Theorem II, we know that $\det(A_{E_0}^{-1/2} A_E A_{E_0}^{-1/2})$ exists. Therefore, using Eq.(4.10), we can perform the following manipulations:

$$\det\left(A_{E_0}^{-1/2} A_E A_{E_0}^{-1/2}\right) = \det\left(1 + A_{E_0}^{-1/2}(A_E - A_{E_0}) A_{E_0}^{-1/2}\right)$$
$$= \det\left(1 + A_{E_0}^{-1/2} \gamma (G_0(E) - G_0(E_0)) \gamma^* A_{E_0}^{-1/2}\right)$$
$$= \det\left(1 + A_{E_0}^{-1/2} \gamma \left(\frac{1}{H_0 - E} - \frac{1}{H_0 - E_0}\right) \gamma^* A_{E_0}^{-1/2}\right)$$
$$= \det\left(1 + (E - E_0) A_{E_0}^{-1/2} \gamma \frac{1}{H_0 - E_0} \frac{1}{H_0 - E} \gamma^* A_{E_0}^{-1/2}\right)$$
$$= \det\left(1 + (E - E_0) \frac{1}{H_0 - E_0} \gamma^* A_{E_0}^{-1} \gamma \frac{1}{H_0 - E}\right)$$
$$= \det\left(1 - \frac{1}{H_0 - E_0} \gamma^* A_{E_0}^{-1} \gamma \frac{1}{H_0 - E_0} \frac{1}{G_0(E_0) - (E - E_0)^{-1}}\right)$$
$$= \det\left(1 + V \frac{1}{G_0(E_0) - (E - E_0)^{-1}}\right) \ .$$

The second to last equality follows from Eq.(4.3). This proves Eq.(4.6) and completes the proof of Lemma 4.2.

## 5. Proof of the Structure Theorem I

The proof of the Structure Theorem I is in two steps. We first decompose $A_E(s,s')$ as

$$A_E(s,s') = -\frac{1}{2\pi}\log\bigl(\sqrt{E}r(s,s')\bigr) + R(\sqrt{E}r(s,s')), \qquad (5.1)$$

where $r(s,s') = |x(s) - x(s')|$. The idea is that the logarithmic term is the most singular one in Eq.(1.2), and that all terms coming from $R_E(s,s') \equiv R(\sqrt{E}r(s,s'))$ are more regular near the origin.

In order to bound $\Lambda A_E$, we bound the contributions from the logarithmic term and from $\Lambda R_E$ separately. The logarithmic term is at the origin of the two contributions $\frac{1}{2} + B + H$ in Eq.(2.4) and $\frac{1}{2}P_0 \log E$ which we add in the estimates of Eq.(2.5). The first piece has been analyzed in detail in [EP] and we will not repeat this analysis here. The reader should observe that the first piece is *independent* of $E$, and that the $E$-dependent terms are trace class.

The operator whose integral kernel is identically equal to 1 is of rank one and maps to the constant functions. Therefore,

$$(2\pi)^{-1}\Lambda\log(\sqrt{E}r) = \tfrac{1}{2}P_0\log E + (2\pi)^{-1}\Lambda\log r. \qquad (5.2)$$

The Structure Theorem I is a consequence of the decomposition

$$-\frac{1}{2\pi}\Lambda\log r = \tfrac{1}{2} + B + H,$$

which we proved in [EP], of the identity (5.2), and of the new estimate

**Theorem 5.1.** *There is a constant c such that for all $E \in \mathcal{E}$, one has the bound*

$$\|\Lambda R_E \Lambda\|_2 \leq c|E|^{3/4+}. \qquad (5.3)$$

Clearly, this shows the second inequality of Eq.(2.5). The first one follows then from Eq.(2.9) and

$$\langle T_E^{(1)}\rangle_{2/3} = \langle T_E^{(1)}\Lambda\Lambda^{-1}\rangle_{2/3} \leq 2^{3/2}\langle T_E^{(1)}\Lambda\rangle_2 \langle\Lambda^{-1}\rangle_1 \leq \text{const.}\|\Lambda R_E\Lambda\|r_2.$$

The proof of the Structure Theorem I is complete.

**Proof of Theorem 5.1.** Since $i\partial_s$ is selfadjoint, we have the following representation for $\Lambda$:

$$\Lambda = \bigl(1 + (i\partial_s)^2\bigr)^{1/2} = |1 + \partial_s| = U(1 + \partial_s),$$

where $U$ is unitary. Similarly, we also have $\Lambda = (1 - \partial_s)U^*$. In view of these identities and the fact that $\langle\cdot\rangle_p$ is a unitary invariant, one has

$$\langle\Lambda R_E\Lambda\rangle_2 = \langle(1 + \partial_s)R_E(1 - \partial_s)\rangle_2.$$

We shall bound this latter quantity. It is useful to introduce $k = \sqrt{E}$. From the definition (5.1), we find

$$R(z) = -\tfrac{1}{4}Y_0(z) + \tfrac{i}{4}J_0(z) + \log(z)/(2\pi) .$$

Below, we shall use some detailed properties of these functions. We note that the kernel of $(1 + \partial_s)R_E(1 - \partial_s)$ is

$$M(s, s') = (1 + \partial_s)(1 + \partial_{s'})R_E(s, s') .$$

Defining $r_s = \partial_s r$, $r_{s'} = \partial_{s'} r$, and $\Psi(z) = R'(z)/z$, one gets

$$M(s, s') = R(kr) + k^2 r\bigl(r_s + r_{s'} + r_{ss'}\bigr)\Psi(kr) + k^2 R''(kr) r_s r_{s'} . \tag{5.4}$$

From the definition of $r$ one finds

$$r_s = \frac{\bigl(x(s) - x(s')\bigr) \cdot \partial_s x(s)}{r} ,$$

$$r_{ss'} = -\frac{1}{r}\bigl(\partial_s x(s) \cdot \partial_{s'} x(s') - r_s r_{s'}\bigr) .$$

We analyze in detail the Green's function in two regions which are defined by

$$D_1 \equiv \{(s, s') : |k| \cdot |x(s) - x(s')| \leq 1\} ,$$
$$D_2 \equiv \{(s, s') : |k| \cdot |x(s) - x(s')| > 1\} .$$

Corresponding to this decomposition, we write

$$M = M_1 + M_2 , \quad M_j(s, s') = M(s, s')\chi(\{(s, s') \in D_j\}) , j = 1, 2 .$$

We shall bound the Hilbert-Schmidt norms of $M_1$ and $M_2$. Throughout, we use the following important inequality which holds for a (connected) standard domain: There is a constant $C > 0$ so that for all $s, s'$ one has

$$C|s - s'| \leq |x(s) - x(s')| \leq |s - s'| . \tag{5.5}$$

In the domain $D_1$, we use the known expansions for the functions $J_0$ and $Y_0$. They are, near $z = 0$,

$$J_0(z) = 1 - \frac{1}{4}z^2 + \mathcal{O}(z^4) ,$$

$$Y_0(z) = \frac{2}{\pi}\log(z)J_0(z) + \hat{Y}_0(z) ,$$

where $J_0$ and $\hat{Y}_0$ are analytic near $z = 0$. It will be useful to write $Y_0$ as

$$Y_0(z) = \frac{2}{\pi}\log(z) + \frac{2}{\pi}\log(z)(J_0(z) - 1) + \hat{Y}_0(z) .$$

Note that the first term in $Y_0$ is generating the logarithmic term of Eq.(5.1), so that only the sum of all other terms contributes to $M$. Using Eq.(5.4), it can be bounded by

$$|M| \leq |R(kr)| + \text{const.}|k|^2|\Psi(kr)| + \text{const.}|k|^2|R''(kr)| . \tag{5.6}$$

Since $R(z) = -\frac{1}{4}Y_0(z) + \frac{i}{4}J_0(z) + \log(z)/(2\pi)$, the expansions near $z = 0$ lead to the bounds

$$\begin{aligned} R(z) &= \mathcal{O}(1)z^2 \log(z) , \\ \Psi(z) &= \mathcal{O}(1)\log(z) , \\ R''(z) &= \mathcal{O}(1)\log(z) . \end{aligned}$$

Therefore, we get from Eq.(5.6),

$$|M_1|(s,s') \leq \text{const.}(1 + |k|^2|\log(kr)|) . \tag{5.7}$$

Thus, the Hilbert-Schmidt norm of $M_1$ is bounded by

$$\|M_1\|_2^2 \leq \text{const.} \int_{|kr|<1} ds\, ds' (1 + |k|^2|\log(kr)|)^2 \leq \mathcal{O}(1)|k|^3 .$$

The last bound follows because $|kr(s,s')| < 1$ implies $|s - s'| < \mathcal{O}(k^{-1})$, by Eq.(5.5).

In the complement of this region, we use that $M_2(s,s')$ is bounded by $\mathcal{O}(k^2)(kr)^{-1/2} + \log(kr)$: this follows again from the explicit representations of the Bessel functions whose derivatives all decay like $z^{-1/2}$ for large $z$. It also has compact support (uniformly in $k$). Therefore, the Hilbert-Schmidt norm of $M_2$ can be bounded by

$$\begin{aligned} \|M_2\|_2^2 &\leq \int_{1<|k(s-s')|<|k|d} ds\, ds' \left|\frac{k^4}{kr(s,s')}\right| + \left|\log(kr(s,s'))\right|^2 \\ &\leq \int_{1<|k(s-s')|<|k|d} ds\, ds' \left|\frac{k^4}{C|s-s'|}\right| + \left|\log(k|s-s'|)\right|^2 \leq \text{const.}|k|^3|\log k| . \end{aligned}$$

The proof of Theorem 5.1 is complete.

**Remark.** It is only this last quantity which leads to the logarithmic corrections of the power laws in $|E|^{3/4}$. We believe that a bound

$$\langle M_2 \rangle_2 \leq \text{const.}|E|^{3/4}$$

should be valid.

**Acknowledgments.** We have profited from helpful discussions with A. Jensen, U. Smilansky and I. Ussishkin. This work has been supported by the Fonds National Suisse.